# Computation of the Activity-on-Node Binary-State Reliability with Uncertainty Components

Wei-Chang Yeh
Department of Industrial Engineering and Engineering Management
National Tsing Hua University
P.O. Box 24-60, Hsinchu, Taiwan 300, R.O.C.
yeh@ieee.org

*Abstract* — Various networks such as cloud computing, water/gas/electricity networks, wireless sensor networks, transportation networks, and 4G/5G networks, have become an integral part of our daily lives. A binary-state network (BN) is often used to model network structures and applications. The BN reliability is the probability that a BN functions continuously; that is, that there is always a simple path connected between a specific pair of nodes. This metric is a popular index for designing, managing, controlling, and evaluating networks. The traditional BN reliability assumes that the reliability of each arc is known in advance. However, this is not always the case. Functioning components operate under different environments; moreover, a network might have newly installed components. Hence, the reliability of these components is not always known. To resolve the aforementioned problems, in which the reliability of some components of a network are uncertain, we introduce the fuzzy concept for the analysis of these components, and propose a new algorithm to solve this uncertainty-component BN reliability problem. The time complexity of the proposed algorithm is analyzed, and the superior performance of the algorithm is demonstrated through examples.

## 1. INTRODUCTION

In binary-state networks (BNs), each unreliable component has only two states. The BN architecture is the most common structure used to model issues in networks that have become an integral part of our daily life, such as networks involving communication [1], network topology design [2], transportation [3, 4], grid/cloud computing [5, 6], data mining [7], Internet of things [8, 9], network resilience problems [10, 11], distribution [12], transition [13], and the joint generation [14]. Hence, an increasing amount of BN applications have recently emerged [15, 16]. BNs thus play an important role in the planning, design, execution, management, and control of all the aforementioned systems [17, 18, 19].

Assume that $G(V, E, \mathbf{D})$ is a BN, where $\mathbf{D}$ is the state distribution of the components, and $V = \{1, 2, \ldots, n\}$ and $E = \{a_1, a_2, \ldots, a_m\}$ are the sets of nodes and arcs, respectively [20, 21]. In the activity-on-arc (AOA) BN, each node is considered perfect, whereas each arc may fail based on a predefined probability in $\mathbf{D}$ [22, 23]. On the contrary, in the activity-on-node (AON) BN, each arc is considered perfect, whereas each node (except the source node 1 and the sink node $n$) may fail based on a predefined probability in $\mathbf{D}$ [24, 25].

Note that a BN can be a mixture of AOA and AON; and a component can be a node, an arc, cable, the Internet, machine, process station, workstation, tool, unit, or worker [26]. In this study, we focus on the AON—and the component is a node in the AON, and an arc in the AOA.

For example, Fig. 1 shows an AOA BN $G(V, E, \mathbf{D})$, where $V = \{1, 2, 3, 4, 5\}$, $E = \{a_1 = e_{1,2}, a_2 = e_{1,3}, a_3 = e_{1,4}, a_4 = e_{2,4}, a_5 = e_{2,5}, a_6 = e_{3,5}\}$. Node 1 is the source node and node 5 is the sink node; the state distribution $\mathbf{D}$ is listed in Table 1.

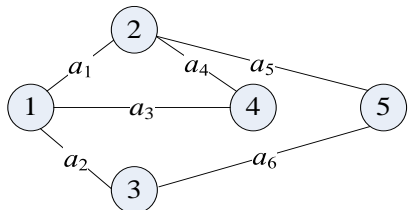

**Figure 1.** Example of an AOA BN.

**Table 1.** State distributions in Fig. 1.

| $a$ | D($a$) |
|---|---|
| $a_1$ | 0.85 |
| $a_2$ | 0.80 |
| $a_3$ | 0.85 |
| $a_4$ | 0.80 |
| $a_5$ | 0.75 |
| $a_6$ | 0.90 |
| $a_7$ | 0.95 |
| $a_8$ | 0.88 |

The traditional reliability, $R$, of BNs is defined as the success probability that there is at least one simple path from nodes 1 to *n*. It is one of the most popular indices for evaluating the performance of real-world BNs. For example: Zhang et al. proposed a energy technology to evaluate the vehicle grid BN reliability [1]. Lin et al. optimized the BN reliability by designing the WSN topology [2]. Aven proposed an algorithm for the transportation BN [3]. Bhavathrathan and Patil also simulated the transportation BN reliability [4]. Yeh and Wei hybrided a binary-code Genetic Algorithm and the simplified swarm optimization (SSO) to optimize the grid BN reliability [6]. Yeh and Lin employed the SSO to optimize the BN reliability in the Internet of Things [8]. Kakadia and Ramirez-Marquez evaluated the telecom BN reliability by considering network resiliency [10]. Laitrakun and Coyle implemented a splitting method to calculate the BN reliability under time constraint in wireless sensor network (WSN) [12]. Ramirez-Marquez solved for the BN reliability using the unscented transformation [13]. As well as, Yeh proposed a squeezed artificial neural network to evaluate the BN reliability [25].

In the traditional BN, the reliability of each component must be known in advance [19, 22] —for example, the reliability $\mathbf{D}(a)$ is already known in Table 1 for all $a \in E$. It is HP-Hard and #P-Hard to calculate the BN reliability [17, 18, 19]. Various algorithms have been proposed to calculate BN reliability and can be categorized into direct and indirect algorithms [27, 28, 29, 30]. The former, for example, the quick binary-addition-tree algorithm (BAT) [31], and binary-decision diagram (BDD) [32] can calculate the BN reliability directly; the latter is separated further into two main steps:

1. Find all simple paths or cuts based on path-based algorithms [2, 5, 8, 9, 21], or cut-based algorithms [20, 22, 23], respectively; and

2. Calculate the BN reliability using inclusion–exclusion technology (IET) [33, 34, 35], or sum–of–disjoint product (SDP) [36, 37, 38, 39].

The direct algorithm is more efficient than the indirect algorithm in calculating the BN reliability because the two main steps in all indirect algorithms are both NP-hard and #P-hard [31].

In addition, regardless of whether we use direct or indirect algorithms, the component is an arc—and the reliability of each component must all be known in advance [26, 27, 30, 35].

In real life, it is not practical or reasonable to assume that the reliability of each component is known precisely beforehand [40, 41, 42]. For example, the Mars Rover is a novel vehicular system, and we cannot know the exact reliability of each component in an environment that is completely different from Earth. Hence, the goal of this study is to overcome the aforementioned obstacle of uncertainty by proposing a new algorithm to deal with uncertain nodes using fuzzy set theory [43, 44] and the binary-addition-tree algorithm (BAT) [11, 26, 45-49].

It is well known that fuzzy set theory can effectively resolve uncertain and imprecise problems. The BAT, first proposed by Yeh [45], can find all possible feasible and required vectors simply based on binary addition. Its current version, quick BAT [31], is more efficient compared to BBD and all other indirect algorithms in calculating BN reliability. Hence, the proposed algorithm is based on BAT combined with fuzzy set theory to solve the proposed uncertainty-component BN reliability problem owing to its importance and practicality.

The remainder of this paper is organized as follows. All acronyms, notations, nomenclatures, and assumptions are introduced in Section 2. A review of AOA BAT, AOA path-based layered-search algorithm (PLSA), and some background material on fuzzy set theory is given in Section 3. The proposed BAT-based algorithm is introduced formally in Section 4, including the preprocessing steps of transferring the linguistic variable via many stages to the unreliability, and the proposed AON PLSA to verify the connectivity of each vector obtained from the proposed BAT. The performance of the proposed algorithm for solving the proposed problem is demonstrated as an example. Finally, Section 5 concludes the study.

## 2. ACRONYMS, NOTATIONS, NOMENCLATURES, AND ASSUMPTIONS

All needed acronyms, notations, assumptions, and nomenclatures are defined and provided here.

## 2.1 Acronyms

BAT: binary-addition-tree algorithm [45]

BN: Binary-state flow network

MP: minimal path

MC: minimal cut

DFS: depth-search-first algorithm

BFS: breadth-search-first algorithm

BBD: binary-decision-diagram

IET: inclusion–exclusion technology

UGFM: universal generating function methodology

PLSA: path-based layered-search algorithm

AOA: activity-on-arc

AON: activity-on-node

AFN: average fuzzy number

FPS: fuzzy possibility score

FFR: fuzzy failure rating

## 2.2 Notations

$/\bullet/$: number of elements in set $\bullet$

$n$: number of nodes

$m$: number of arcs

$a_k$: $k$th undirected arc

$V$: set of nodes $V = \{1, 2, \dots, n\}$ and $|V| = n$

$E$: set of arcs $E = \{a_1, a_2, \dots, a_m\}$ and $|E| = m$

$e_{i,j}$: $e_{i,j} = a_k \in E$, $i, j \in V$, and for one and only one $k$

$Pr(\bullet)$: occurrence probability of $\bullet$

**D**: state distributions listing all components' states and their probabilities, e.g., Table 1 is the arc state distribution of Fig. 1.

$G(V, E)$: A graph created by $V$ and $E$. For example, Fig. 1 shows a network constituted by $E = \{a_1, a_2, \ldots, a_6\}$ and $V = \{1, 2, 3, 4, 5\}$.

$G(V, E, \mathbf{D})$: A network constituted by $G(V, E)$ and **D**. For example, $G(V, E)$ shown in Fig. 1 and **D** in Table 1 form a network.

$X$: binary state vector; its $k$th coordinate is the state of the $k$th component

$x_k$: state of the $k$th coordinate in $X$

$Pr(\bullet)$: occurrence probability of component $\bullet$

$Pr(X)$: occurrence probability of vector $X$

$R$    Reliability of a BN

$G(X)$: $G(X) = G(V, \{a \in E \mid \text{each arc } a \text{ with } X(a) > 0\}, \mathbf{D})$

$X \ll Y$: if vector $X$ is obtained before $Y$ in the BAT.

### 2.3 Nomenclatures

MP/MC: An MP/MC is an arc subset; each of its proper subsets is not an MP/MC [17,19,25]. For example, $\{a_1, a_3, a_6\}$ is an MC, and $\{a_3, a_4, a_5\}$ is an MP in Fig. 1.

Uncertainty component: The reliability of the component is uncertain.

Crisp component: The reliability of the component is certain.

$R$: Success probability of at least a directed path from nodes 1 to $n$

### 2.4 Assumptions

1. The networks have no loops or parallel arcs.

2. Each node and arc are perfectly reliable in AOA BN and AON BN, respectively.

3. The source node and the sink node, that is, nodes 1 and $n$, are perfectly reliable.

4. Each arc is undirected.

5. The states of the components are statistically independent, and its state follows **D** to be zero or one.

## 3. OVERVIEW OF BAT AND FUZZY SET

The proposed AON BAT is modified from the traditional AOA (binary-state) BAT [45] to find each vector, say *X*. It is also used to verify whether *X* is connected using PLSA [45] after transferring each uncertainty component with fuzzy and incomplete information to a crisp component to calculate the uncertainty-component binary-state AON reliability. Hence, the traditional AOA BAT, AOA PLSA, and fundamental fuzzy set-related parts are discussed in this section.

### 3.1 AOA BAT

(AOA) BAT was first proposed by Yeh [45]; based on binary addition, it is used to solve AOA binary-state network reliability problems. Owing to its simplicity, efficiency, and memory saving capability, BAT has been implemented in various areas; it has recently become a new search method for finding all possible solutions or vectors. For example, BATs have been implemented to solve rework problems [26], network reliability problems [31, 45, 48, 49], wildfire spread prediction [46], and forecasting of computer virus propagation [47].

From the results of numerical experiments evaluating running time and computer memory, BAT [45] was found to outperform depth-search-first algorithms (DFS)—for example, the quick IET [34], which is the best-known algorithm for calculating MFN reliability; the breadth-search-first algorithm (BFS), the universal generating function methodology (UGFM) [19], which is the main algorithm for solving information network reliability problems [50]; and binary-decision diagram (BBD) [32], which outperforms other algorithms in binary-state network reliability problems [17, 18].

The purpose of the original BAT is to search all possible vectors for related AOA problems [45]. BAT starts from a vector zero, say *X*, and adds one to *X* to update *X* iteratively. Let us briefly explain the basic principles of BAT. The entire original BAT is processed in a backward manner from the last coordinate of $X$ to the first coordinate [45]. Let $x_k$ be the current coordinate. If $x_k = 0$, $x_k$ is updated by letting $x_k = 1$. A new vector is formed, and the current coordinate is reset to $m$—e.g., (1, 0, 0) is updated to (1, 0, 1). If $x_k = 1$, $x_k$ is updated by letting $x_k = 0$. The process is then moved to the $(k-1)$th coordinate to repeat the aforementioned procedure until the new current coordinate is zero—e.g., (1, 0, 1) is updated to (1, 1, 0).

The original AOA BAT pseudocode is as follows [45]:

**Algorithm:** AOA BAT

**Input:**   A $m$-tuple zero vector $X$.

**Output:**   All related vectors of the related problem.

**STEP B0.**   Let SUM = 0, $X = \mathbf{0}$, and $k = m$.

**STEP B1.**   Assign SUM = SUM − 1, $X(a_k) = 0$, and go to STEP B4 if $X(a_k) = 1$.

**STEP B2.**   Assign SUM = SUM + 1 and $X(a_k) = 1$.

**STEP B3.**   Assign $k = k − 1$ and go to STEP B1 if $k > 1$.

**STEP B4.**   If SUM = $m$, halt; otherwise, let $k = m$ and go to STEP B1.

From the aforementioned 5 lines of pseudocode, it can be observed that BAT is simple to program, efficient to run, easy to understand, economical in RAM, and suitable for make-to-fit [11, 26, 45-49].

The number of all obtained vectors is $2^m$ for an $m$-tuple vector zero in STEP B0. The time complexity of BAT is $O(2^{m+1})$, as proved in [49]. Furthermore, the total computer memory is only $O(m)$ because the current vector $X$ is used repeatedly, and a new updated vector needs to compute only its related values or predefined functions [45].

The following table shows how the AOA BAT is implemented to find all the state vectors in Fig. 1. Because $|E| = 6$ in Fig. 1, there are $2^6 = 64$ vectors at the end. To recognize the sequence of the obtained vectors easily, the notation $X_i$ is used in Table 2. However, in the actual BAT procedure, there is always one vector *X* in the complete process; there is no need to discriminate the sequence of the obtained *X*.

**Table 2.** Complete results obtained from the BAT on Fig. 1.

| i | $X_i$ | i | $X_i$ | i | $X_i$ | i | $X_i$ |
|---|---|---|---|---|---|---|---|
| 1 | (0, 0, 0, 0, 0, 0) | 17 | (0, 0, 0, 0, 1, 0) | 33 | (0, 0, 0, 0, 0, 1) | 49 | (0, 0, 0, 0, 1, 1) |
| 2 | (1, 0, 0, 0, 0, 0) | 18 | (1, 0, 0, 0, 1, 0) | 34 | (1, 0, 0, 0, 0, 1) | 50 | (1, 0, 0, 0, 1, 1) |
| 3 | (0, 1, 0, 0, 0, 0) | 19 | (0, 1, 0, 0, 1, 0) | 35 | (0, 1, 0, 0, 0, 1) | 51 | (0, 1, 0, 0, 1, 1) |
| 4 | (1, 1, 0, 0, 0, 0) | 20 | (1, 1, 0, 0, 1, 0) | 36 | (1, 1, 0, 0, 0, 1) | 52 | (1, 1, 0, 0, 1, 1) |
| 5 | (0, 0, 1, 0, 0, 0) | 21 | (0, 0, 1, 0, 1, 0) | 37 | (0, 0, 1, 0, 0, 1) | 53 | (0, 0, 1, 0, 1, 1) |
| 6 | (1, 0, 1, 0, 0, 0) | 22 | (1, 0, 1, 0, 1, 0) | 38 | (1, 0, 1, 0, 0, 1) | 54 | (1, 0, 1, 0, 1, 1) |
| 7 | (0, 1, 1, 0, 0, 0) | 23 | (0, 1, 1, 0, 1, 0) | 39 | (0, 1, 1, 0, 0, 1) | 55 | (0, 1, 1, 0, 1, 1) |
| 8 | (1, 1, 1, 0, 0, 0) | 24 | (1, 1, 1, 0, 1, 0) | 40 | (1, 1, 1, 0, 0, 1) | 56 | (1, 1, 1, 0, 1, 1) |
| 9 | (0, 0, 0, 1, 0, 0) | 25 | (0, 0, 0, 1, 1, 0) | 41 | (0, 0, 0, 1, 0, 1) | 57 | (0, 0, 0, 1, 1, 1) |
| 10 | (1, 0, 0, 1, 0, 0) | 26 | (1, 0, 0, 1, 1, 0) | 42 | (1, 0, 0, 1, 0, 1) | 58 | (1, 0, 0, 1, 1, 1) |
| 11 | (0, 1, 0, 1, 0, 0) | 27 | (0, 1, 0, 1, 1, 0) | 43 | (0, 1, 0, 1, 0, 1) | 59 | (0, 1, 0, 1, 1, 1) |
| 12 | (1, 1, 0, 1, 0, 0) | 28 | (1, 1, 0, 1, 1, 0) | 44 | (1, 1, 0, 1, 0, 1) | 60 | (1, 1, 0, 1, 1, 1) |
| 13 | (0, 0, 1, 1, 0, 0) | 29 | (0, 0, 1, 1, 1, 0) | 45 | (0, 0, 1, 1, 0, 1) | 61 | (0, 0, 1, 1, 1, 1) |
| 14 | (1, 0, 1, 1, 0, 0) | 30 | (1, 0, 1, 1, 1, 0) | 46 | (1, 0, 1, 1, 0, 1) | 62 | (1, 0, 1, 1, 1, 1) |
| 15 | (0, 1, 1, 1, 0, 0) | 31 | (0, 1, 1, 1, 1, 0) | 47 | (0, 1, 1, 1, 0, 1) | 63 | (0, 1, 1, 1, 1, 1) |
| 16 | (1, 1, 1, 1, 0, 0) | 32 | (1, 1, 1, 1, 1, 0) | 48 | (1, 1, 1, 1, 0, 1) | 64 | (1, 1, 1, 1, 1, 1) |

### 3.2 AOA PLSA

For solving network reliability problems in the traditional BAT, the connectivity of each obtained vector found in the BAT procedure is verified using AOA PLSA [45]. Consequently, the summation of the probabilities of all connected vectors is related to network reliability.

Each vector, say *X*, obtained from BAT is a subgraph $G(V, X)$ of $G(V, E)$. For example, the subgraph corresponding to $X_{28} = (1, 1, 0, 1, 1, 0)$ is depicted in Fig. 2.

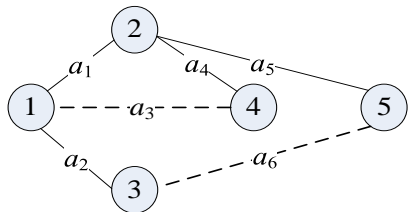

**Figure 2.** Subgraph $X_{28} = (1, 1, 0, 1, 1, 0)$.

The layered-search algorithm (LSA) was first proposed in [51] and revised to PLSA to verify the connectivity of each vector obtained from BAT [45].

To apply the PLSA to test whether $X$ is connected i.e., whether there is a simple path from nodes 1 to $n$ in $G(V, X)$, an empty queue is initialized by adding node 1 to the queue. This queue stores the nodes that are adjacent to any node in the queue by at least one arc repeatedly. If node $n$ is included, a simple path is found from nodes 1 to $n$—that is, such a vector is connected. Otherwise, $X$ is a disconnected vector. The pseudocode of the AOA PLSA is as follows.

**Algorithm:** AOA PLSA.

**Input:** A vector $X$ obtained from BAT.

**Output:** Whether $X$ is connected or disconnected

**STEP P0.** Let $V^* = Q_0 = \{1\}$ and $i = 1$.

**STEP P1.** Let $Q_i = \{ v \notin V^* \mid \text{for all } e_{u,v} \in E, \text{ and } u \in V^*, \text{i.e., } X(e_{u,v}) = 1\}$.

**STEP P2.** If $n \in Q_i$, $X$ is connected and the algorithm is halted.

**STEP P3.** If $Q_i = \emptyset$, $X$ is disconnected and the algorithm is halted.

**STEP P4.** Let $V^* = V^* \cup Q_i$, $i = i + 1$, and go to STEP P1.

The time complexity of the PLSA to verify whether a vector is connected is $O(n)$. For example, in Fig. 2, the PLSA determines $X = (1, 1, 0, 1, 1, 0)$ to be connected, and the procedure is listed in Table 3.

**Table 3.** Process from the proposed PLSA for Fig. 2.

| $i$ | $Q_i$ | $Q_{i+1}$ | $V^*$ | Remark |
|---|---|---|---|---|
| 0 | {1} | {2, 3} | {1, 2, 3} | |
| 1 | {2, 3} | {5} | | $X$ is connected |

### 3.3 Fuzzy Set Theory

The fuzzy set theory introduced by Zadeh has become a common technique for handling uncertain problems with imprecise information [43]. The basic principle of fuzzy set theory is briefly presented in this section.

### 3.3.1 Fuzzy Number

Let $x$ be a real number allocated to $\underline{A}$, and $\underline{A} \subset$ be the universe set $U$. The fuzzy set $\underline{A}$ is defined as follows [43]:

$$\underline{A} = \{(x, u_{\underline{A}}(x)) \mid x \in X\} \tag{1}$$

where $u_{\underline{A}}(x): X \rightarrow [0, 1]$ denotes the membership function of fuzzy set $A$.

There are many forms of fuzzy numbers that represent fuzzy sets. In this study, triangular fuzzy numbers are applied. A triangular fuzzy number $\underline{A} = (a, b, c)$, where $a \leq b \leq c$, is defined in Eq. (2); it is depicted in Figure 3 [43]:

$$u_{\underline{A}}(x) = \begin{cases} (x-a)/(b-a), & a \leq x \leq b, \\ (x-c)/(b-c), & b \leq x \leq c, \\ 0, & \text{otherwise}, \end{cases} \tag{2}$$

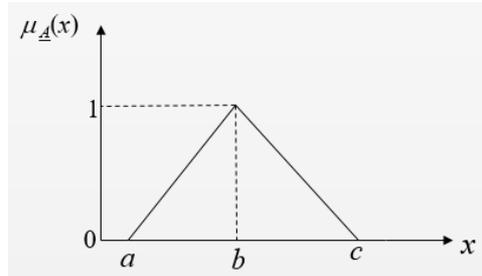

**Figure 3**. Example of a triangular fuzzy number

Let $\underline{A}_1 = (a_1, b_1, c_1)$ and $\underline{A}_2 = (a_2, b_2, c_2)$ be two triangular fuzzy numbers. The fuzzy addition, subtraction, multiplication, and division of triangular fuzzy numbers are as shown in Eqs. (3)–(6), respectively [43].

$$\underline{A}_1 \oplus \underline{A}_2 = (a_1 + a_2, b_1 + b_2, c_1 + c_2) \tag{3}$$

$$\underline{A}_1 \ominus \underline{A}_2 = (a_1 - c_2, b_1 - b_2, c_1 - a_2) \tag{4}$$

$$\underline{A}_1 \otimes \underline{A}_2 = (a_1 \times a_2, b_1 \times b_2, c_1 \times c_2) \tag{5}$$

$$\underline{A}_1 \odot \underline{A}_2 = (a_1/c_2, b_1/b_2, c_1/a_2) \tag{6}$$

The α-cut is another common way to represent the fuzzy set; it is also an important approach for transforming a fuzzy membership function into a crisp subset [43]. The α-cut is a horizontal representation of fuzzy sets. Let $A_{α,UB}$ and $A_{α,LB}$ be the upper and lower bounds of the α-cut, respectively. The α-cuts of fuzzy set $A$ can be defined as in Eq. (3) [43]:

$$A_α = \{(x, u_A(x) \geq α) \mid x \in U\} = [A_{α,LB}(α)\ A_{α,UB}(α)], \tag{7}$$

For example, as shown in Fig. 4, the α-cut for the triangular fuzzy number $A = (a, b, c)$ is

$$A_α = [α(b - a) + a, α(b - c) + c]. \tag{8}$$

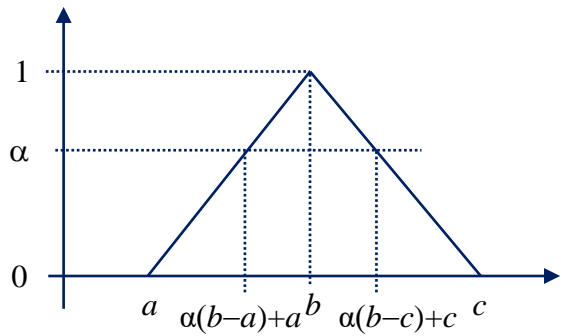

**Figure 4**. α-cut and a triangular fuzzy number

### 3.3.2 Linguistic Variable

It is natural to use linguistic variables to estimate, rate, and judge subjective events. In this study, each uncertainty component had to be rated by experts using linguistic variables. This is shown in the first column of Table 4. To fuzzify the experts' ratings for measuring the reliability of each uncertainty component, the linguistic variables presented in Table 4 are transferred to their approximate reasoning of triangular fuzzy numbers, as shown in the second column of Table 4 [43, 44].

**Table 4**. Linguistic variables and their fuzzy numbers.

| Linguistic Variables | Triangular Fuzzy Numbers |
|---|---|
| Very Low (VL) | (0.0, 0.0, 0.1) |
| Low (L) | (0.0, 0.1, 0.3) |
| Fairly Low (FL) | (0.1, 0.3, 0.5) |
| Medium (M) | (0.3, 0.5, 0.7) |
| Fairly High (FH) | (0.5, 0.7, 0.9) |
| High (H) | (0.7, 0.9, 1.0) |
| Very High (VH) | (0.9, 1.0, 1.0) |

For example, assume that components 3, 4, and 5 are uncertainty nodes, and six experts are asked to rate these components using linguistic variables. The results of these linguistic variables are listed in Table 5.

**Table 5**. Linguistic variables of six experts.

| component | Linguistic Variables |
|---|---|
| 3 | VL, L, L, VL, L, L |
| 4 | VL, L, H, VH, L, H |
| 5 | VH, H, L, VH, H, L |

### 3.3.3 Average Fuzzy Number

After obtaining linguistic variables for these components with uncertain information, we need to aggregate the experts' rates; that is, combine the linguistic variables into one fuzzy number called the average fuzzy number (AFN)—by averaging the fuzzy number of all experts' linguistic variables based on the $\alpha$-cut [44].

Assume that $A_{i,j} = (a_{i,j}, b_{i,j}, c_{i,j})$ is the linguistic variable provided by expert $j$ for uncertainty component $i$. The average fuzzy number of the uncertainty component $i$ is defined as

$$A_i = (A_{i,1} \oplus A_{i,2} \oplus \ldots \oplus A_{i,h})/h \tag{9}$$

where $h$ is the number of experts.

For example, six experts rated the uncertainty component 3: VL, L, L, VL, L, and L. Based on Eq. (3), Eq. (9) and Tables 4 and 5, we obtain the AFN for the uncertainty component 3:

$$A_3 = [2(0.0, 0.0, 0.1) + 4(0.0, 0.1, 0.3)]/6 = (0.0, 0.4/6, 1.4/6). \tag{10}$$

### 3.3.4 Defuzzification to Convert AFN to FPS

The AFN is still a fuzzy number and requires further defuzzification to convert it to a crisp number called the fuzzy possibility score (FPS)—which represents the experts' belief of the most likely score of an event occurring. The method of transferring AFN to FPS is based on the left and right fuzzy ranking method proposed in [52]. To implement this method, we need to define the fuzzy maximizing and minimizing sets as:

$$f_{max}(x) = \begin{cases} x & x \in [0,1] \\ 0 & \text{otherwise} \end{cases} \quad (11)$$

$$f_{min}(x) = \begin{cases} 1-x & x \in [0,1] \\ 0 & \text{otherwise} \end{cases}. \quad (12)$$

Based on the α-cut, we have FPS$_R(k)$ and FPS$_L(k)$ for uncertainty component $k$ [52]:

$$\text{FPS}_R(k) = \sup_x [f_k(x) \wedge f_{max}(x)] = \alpha, \text{ where } \alpha = \underset{\sim}{A}_{\alpha,\text{UB}}(\alpha). \quad (13)$$

$$\text{FPS}_L(k) = \sup_x [f_k(x) \wedge f_{min}(x)] = \alpha, \text{ where } (1-\alpha) = \underset{\sim}{A}_{\alpha,\text{LB}}(\alpha), \quad (14)$$

After solving α and bringing α into the following equation, we can obtain the FPS for the uncertainty component $k$ [52]:

$$\text{FPS}(k) = |\text{FPS}_R(k) + 1 - \text{FPS}_L(k)|/2. \quad (15)$$

For example, the α-cut for $\underset{\sim}{A}_3 = (0.0, 0.4/6, 1.4/6)$ is

$$\underset{\sim}{A}_{3,\alpha} = [\alpha(0.4/6 - 0) + 0, \alpha(0.4/6 - 1.4/6) + 1.4/6] = [0.4\alpha/6, 1.4/6 - \alpha/6]. \quad (16)$$

After solving Eq. (13) and Eq. (14), we have

$$\alpha = \underset{\sim}{A}_{3,\alpha,\text{UB}}(\alpha) \Rightarrow \alpha = 1.4/6 - \alpha/6 \Rightarrow \alpha = \text{FPS}_R(3) = 0.2 \quad (17)$$

$$(1-\alpha) = \underset{\sim}{A}_{3,\alpha,\text{LB}}(\alpha) \Rightarrow (1-\alpha) = 0.4\alpha/6 \Rightarrow \alpha = \text{FPS}_L(3) = 0.9375. \quad (18)$$

Hence,

$$\text{FPS}(3) = |\text{FPS}_R(3) + 1 - \text{FPS}_L(3)|/2 = 0.86875. \quad (19)$$

### 3.3.5 Further conversion of FPS to FFR

To confirm the affinity between the non-fuzzy failure rate of hardware and experts' FPS, the FPS must be converted into fuzzy failure rates (FFRs), which is an error rate [53]:

$$\text{FFR} = (\text{frequency of error}) / (\text{total chance that an event has an error}). \quad (20)$$

$$= \begin{cases} 10^{-k} & \text{FPS} \neq 0 \\ 0 & \text{otherwise} \end{cases}, \quad (21)$$

where

$$k = |\ (1 - \text{FPS})\ /\ \text{PFS}\ |^{1/3} \times 2.301. \tag{22}$$

The FFR of an uncertainty component, say $i$, is treated as the unreliability; that is, $R(i) = 1 - \text{FFR}(i)$. For example, because

$$k = |\ (1 - 0.86875)/\ 0.86875\ |^{1/3} \times 2.301 = 1.225514. \tag{23}$$

We have the FFR of uncertainty component 3:

$$\text{FFR}(3) = 10^{-k} = 10^{-1.225514} = 0.059496, \tag{24}$$

and

$$R(3) = 1 - \text{FFR}(3) = 0.940504. \tag{25}$$

## 4. PROPOSED AON BAT-BASED ALGORITHM

The proposed AON BAT-based algorithm for computing the reliability of a binary-state network with uncertainty nodes is presented in this section. The preprocessing steps for transferring uncertainty nodes into crisp nodes are described in Section 4.1. Moreover, the first AON BAT is proposed by modifying the traditional AOA BAT in Section 4.2. An example of the proposed AON BAT is presented in Section 4.3.

### 4.1 Preprocessing Operations: Transfer of Uncertainty Components into Crisp Components

The preprocessing steps of the proposed algorithm include dealing with uncertainty nodes whose reliabilities are unknown. The basic concept of transferring uncertainty nodes into crisp nodes is based on fuzzy set theory [43], as mentioned in Section 3.2. The pseudocode for the proposed preprocessing operation is as follows [44]:

**Algorithm:** Preprocess Uncertainty Components

**Input:** An AON BN $G(V, E, \mathbf{D})$ with uncertainty nodes.

**Output:** Convert each uncertainty node into a crisp node.

**STEP P0.** Each expert rates the uncertainty nodes using linguistic variables.

**STEP P1.** Transfer each linguistic variable into a fuzzy number.

**STEP P2.** Average experts' fuzzy numbers and calculate their related α-cuts for each uncertainty node to have an AFN [44].

**STEP P3.** Convert AFN to FPS with defuzzification [52].

**STEP P4.** Convert FPS to FFR [53] —which is the unreliability.

Taking Table 6 as an example of which nodes 3–5 are uncertainty nodes, in STEP P0, experts first rate these uncertainty nodes using linguistic variables, as shown in Table 4 and the second column of Table 6. Then, STEP P2 calculates the average fuzzy number (AFN) based on Table 4 and STEP P1, as shown in the third column of Table 6.

STEP P2 mainly focused on aggregating experts' ratings into an AFN. Subsequently, each α-cut of the related AFN is also required to be calculated—as shown in the last column of Table 6.

Table 6. Example of STEPs P0, P1, and P2.

| Node $i$ | Linguistic Variables | Average Fuzzy Number | α-cut |
|---|---|---|---|
| 3 | VL, L, L, VL, L, L | (0, 0.4/6, 1.4/6) | [0.4α/6, 1.4/6−α/6] |
| 4 | VL, L, H, VH, L, H | (2.3/6, 3/6, 3.7/6) | [0.7α/6+2.3/6, 3.7/6−0.7α/6] |
| 5 | VH, H, L, VH, H, L | (3.2/6, 4/6, 4.6/6) | [0.8α/6+3.2/6, 4.6/6−0.1α] |

After many fuzzy numbers are aggregated into an AFN and the related α-cut, the next step is to defuzzify the values. Based on the left and right fuzzy ranking method proposed in [52], STEP P3 defuzzifies AFN to FPS, as shown in columns 2 to 4 of Table 7. The last step, that is, STEP P4, transfers APS to FFR based on Eq. (21), as listed in the last two columns of Table 7. Consequently, the FFR is treated as the unreliability of the related uncertainty node.

Table 7. Example of STEPs P3 and P4.

| Node $i$ | $FPS_L(i)$ | $FPS_R(i)$ | $FPS(i)$ | $k$ | $FFR(i)$ | $R(i)$ |
|---|---|---|---|---|---|---|
| 3 | 0.2 | 0.9375 | 0.86875 | 1.225514 | 0.059496 | 0.940504 |
| 4 | 0.552239 | 0.552239 | 0.5 | 2.301 | 0.005 | 0.995000 |
| 5 | 0.411764 | 0.696970 | 0.642603 | 1.892283 | 0.012815 | 0.987185 |

## 4.2 AON BAT

The proposed AON BAT is based on the traditional AOA BAT proposed in [45]. In the proposed AON BAT, the values of the first and last coordinates, that is, the source node and the sink node, are always 1 because the BN fails if any one of these two nodes fails.

Unlike the traditional AOA BAT that is implemented backward from the last to the first coordinate, the proposed AON BAT is implemented forward—that is, from the first to the second last node. Note that the states of the first and last nodes are always one and do not need to be changed.

Moreover, in the traditional AOA BAT, there is a variable "SUM," to count the number of coordinates—of which the values are one—to decide when to stop the BAT procedure. To improve the efficiency, the proposed AON BAT adopts another method by checking whether the current coordinate is the last coordinate; that is, the coordinate $n$. If the answer is yes, then the algorithm is stopped.

The pseudocode of the proposed AON BAT is presented in the following.

**Algorithm: AON BAT**

**Input:** A $n$-tuple vector $X$.

**Output:** All solutions in vector form are found in the related AON problems.

**STEP N0.** Let $X = \mathbf{0}$, except that both the first and last are 1s, and $k = 2$.

**STEP N1.** If $X(a_k) = 0$, let $X(a_k) = 1$, $k = 1$, a new $X$ is found, and go to STEP N1.

**STEP N2.** Let $X(a_k) = 0$.

**STEP N3.** If $k < n$, let $k = k + 1$ and go to STEP N1. Otherwise, halt.

The number of all obtained vectors from the proposed AON BAT is $2^{n-2}$ because the 1st and $n$th coordinates are unchanged. In addition, the time complexity of the traditional AOA BAT is $O(2^{m+1})$ with the space complexity being $O(m)$ [49], whereas the time complexity and space complexity of the AON BAT are $O(2^{n-2+1}) = O(2^{n-1})$ and $O(n-1)$, respectively. Moreover, to calculate the AON BN reliability, an additional step is needed.

**STEP N4.** If $X$ is connected, let $R = R + \Pr(X)$. Go to STEP N1.

Correspondingly, STEP N1 needs to be revised as follows:

**STEP N1.** If $X(a_k) = 0$, let $X(a_k) = 1$, $k = 1$, a new $X$ is found, and go to STEP N4.

### 4.3 AON PLSA

The proposed AON PLSA is based on the traditional AOA PLSA proposed in [45, 51] to verify the connectivity of a vector obtained from the proposed AON BAT. The basic idea of the proposed AON PLSA is very similar to that of the AOA PLSA. The pseudocode of the proposed AON PLSA is similar to that of the AOA PLSA and is listed as follows.

**Algorithm:** AOA PLSA.

**Input:** A vector $X$ obtained from BAT.

**Output:** Whether $X$ is connected or disconnected

**STEP L0.** Let $V^* = Q_0 = \{1\}$ and $i = 1$.

**STEP L1.** Let $Q_i = \{ v \notin V^* \mid \text{for all } e_{u,v} \in E, X(v) = 1, \text{ and } u \in V^* \}$.

**STEP L2.** If $n \in Q_i$, $X$ is connected, the algorithm is halted.

**STEP L3.** If $Q_i = \emptyset$, $X$ is disconnected, the algorithm is halted.

**STEP L4.** Let $V^* = V^* \cup Q_i$, $i = i + 1$; then, go to STEP L1.

The only difference between the proposed AON PLSA and AOA PLSA is the second step. Hence, the time complexity of the proposed AON PLSA is $O(n)$ [45].

### 4.4 Example

The general procedure of the proposed AON BAT-based algorithm is best demonstrated with examples. All types of networks are NP-hard and #P-hard. To reduce the complexity, a middle-sized AON BN, as shown in Fig. 5, was implemented to better demonstrate how the proposed algorithm solves the AON BN reliability with the uncertainty components problem.

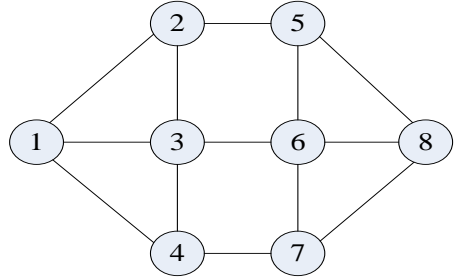

**Figure 5.** Example AON BN

The information of nodes 3, 4, and 5 are unknown, and their ratings, that is, linguistic variables, are provided by six experts; the reliability of nodes 2, 6, and 7 are listed in Table 8. Note that nodes 1 and 8 are perfectly reliable.

**Table 8.** State distributions in Fig. 5.

| $a$ | D($a$) | Linguistic Variables |
|---|---|---|
| $a_2$ | 0.80 | |
| $a_3$ | unknown | VL, L, L, VL, L, L |
| $a_4$ | unknown | VL, L, H, VH, L, H |
| $a_5$ | unknown | VH, H, L, VH, H, L |
| $a_6$ | 0.90 | |
| $a_7$ | 0.88 | |

After the preprocessing steps provided in Section 4.1, all uncertainty components are transferred to crisp components, as shown in Tables 6 and 7. The new state distribution after transformation is as follows:

**Table 9.** New state distributions of Fig. 5.

| $a$ | D($a$) |
|---|---|
| $a_2$ | 0.80 |
| $a_3$ | 0.940504 |
| $a_4$ | 0.995000 |
| $a_5$ | 0.987185 |
| $a_6$ | 0.90 |
| $a_7$ | 0.88 |

**Table 10.** Result obtained from the proposed AON BAT.

| $i$ | $X_i$ | Pr($x_2$) | Pr($x_3$) | Pr($x_4$) | Pr($x_5$) | Pr($x_6$) | Pr($x_7$) | Pr($X_i$) |
|---|---|---|---|---|---|---|---|---|
| 1 | (1, 0, 0, 0, 0, 0, 0, 1) | 0.2 | 0.059496 | 0.005 | 0.012815 | 0.1 | 0.12 | |
| 2 | (1, 1, 0, 0, 0, 0, 0, 1) | 0.8 | 0.059496 | 0.005 | 0.012815 | 0.1 | 0.12 | |
| 3 | (1, 0, 1, 0, 0, 0, 0, 1) | 0.2 | 0.940504 | 0.005 | 0.012815 | 0.1 | 0.12 | |
| 4 | (1, 1, 1, 0, 0, 0, 0, 1) | 0.8 | 0.940504 | 0.005 | 0.012815 | 0.1 | 0.12 | |
| 5 | (1, 0, 0, 1, 0, 0, 0, 1) | 0.2 | 0.059496 | 0.995 | 0.012815 | 0.1 | 0.12 | |
| 6 | (1, 1, 0, 1, 0, 0, 0, 1) | 0.8 | 0.059496 | 0.995 | 0.012815 | 0.1 | 0.12 | |
| 7 | (1, 0, 1, 1, 0, 0, 0, 1) | 0.2 | 0.940504 | 0.995 | 0.012815 | 0.1 | 0.12 | |
| 8 | (1, 1, 1, 1, 0, 0, 0, 1) | 0.8 | 0.940504 | 0.995 | 0.012815 | 0.1 | 0.12 | |
| 9 | (1, 0, 0, 0, 1, 0, 0, 1) | 0.2 | 0.059496 | 0.005 | 0.987185 | 0.1 | 0.12 | |
| 10 | (1, 1, 0, 0, 1, 0, 0, 1) | 0.8 | 0.059496 | 0.005 | 0.987185 | 0.1 | 0.12 | 2.82E-06 |
| 11 | (1, 0, 1, 0, 1, 0, 0, 1) | 0.2 | 0.940504 | 0.005 | 0.987185 | 0.1 | 0.12 | |
| 12 | (1, 1, 1, 0, 1, 0, 0, 1) | 0.8 | 0.940504 | 0.005 | 0.987185 | 0.1 | 0.12 | 4.46E-05 |
| 13 | (1, 0, 0, 1, 1, 0, 0, 1) | 0.2 | 0.059496 | 0.995 | 0.987185 | 0.1 | 0.12 | |
| 14 | (1, 1, 0, 1, 1, 0, 0, 1) | 0.8 | 0.059496 | 0.995 | 0.987185 | 0.1 | 0.12 | 0.000561 |
| 15 | (1, 0, 1, 1, 1, 0, 0, 1) | 0.2 | 0.940504 | 0.995 | 0.987185 | 0.1 | 0.12 | |
| 16 | (1, 1, 1, 1, 1, 0, 0, 1) | 0.8 | 0.940504 | 0.995 | 0.987185 | 0.1 | 0.12 | 0.008869 |
| 17 | (1, 0, 0, 0, 0, 1, 0, 1) | 0.2 | 0.059496 | 0.005 | 0.012815 | 0.9 | 0.12 | |
| 18 | (1, 1, 0, 0, 0, 1, 0, 1) | 0.8 | 0.059496 | 0.005 | 0.012815 | 0.9 | 0.12 | |
| 19 | (1, 0, 1, 0, 0, 1, 0, 1) | 0.2 | 0.940504 | 0.005 | 0.012815 | 0.9 | 0.12 | 1.3E-06 |
| 20 | (1, 1, 1, 0, 0, 1, 0, 1) | 0.8 | 0.940504 | 0.005 | 0.012815 | 0.9 | 0.12 | 5.21E-06 |
| 21 | (1, 0, 0, 1, 0, 1, 0, 1) | 0.2 | 0.059496 | 0.995 | 0.012815 | 0.9 | 0.12 | |
| 22 | (1, 1, 0, 1, 0, 1, 0, 1) | 0.8 | 0.059496 | 0.995 | 0.012815 | 0.9 | 0.12 | |
| 23 | (1, 0, 1, 1, 0, 1, 0, 1) | 0.2 | 0.940504 | 0.995 | 0.012815 | 0.9 | 0.12 | 0.000259 |
| 24 | (1, 1, 1, 1, 0, 1, 0, 1) | 0.8 | 0.940504 | 0.995 | 0.012815 | 0.9 | 0.12 | 0.001036 |
| 25 | (1, 0, 0, 0, 1, 1, 0, 1) | 0.2 | 0.059496 | 0.005 | 0.987185 | 0.9 | 0.12 | |
| 26 | (1, 1, 0, 0, 1, 1, 0, 1) | 0.8 | 0.059496 | 0.005 | 0.987185 | 0.9 | 0.12 | 2.54E-05 |
| 27 | (1, 0, 1, 0, 1, 1, 0, 1) | 0.2 | 0.940504 | 0.005 | 0.987185 | 0.9 | 0.12 | 0.0001 |
| 28 | (1, 1, 1, 0, 1, 1, 0, 1) | 0.8 | 0.940504 | 0.005 | 0.987185 | 0.9 | 0.12 | 0.000401 |
| 29 | (1, 0, 0, 1, 1, 1, 0, 1) | 0.2 | 0.059496 | 0.995 | 0.987185 | 0.9 | 0.12 | |
| 30 | (1, 1, 0, 1, 1, 1, 0, 1) | 0.8 | 0.059496 | 0.995 | 0.987185 | 0.9 | 0.12 | 0.005049 |
| 31 | (1, 0, 1, 1, 1, 1, 0, 1) | 0.2 | 0.940504 | 0.995 | 0.987185 | 0.9 | 0.12 | 0.019954 |
| 32 | (1, 1, 1, 1, 1, 1, 0, 1) | 0.8 | 0.940504 | 0.995 | 0.987185 | 0.9 | 0.12 | 0.079817 |
| 33 | (1, 0, 0, 0, 0, 0, 1, 1) | 0.2 | 0.059496 | 0.005 | 0.012815 | 0.1 | 0.88 | |
| 34 | (1, 1, 0, 0, 0, 0, 1, 1) | 0.8 | 0.059496 | 0.005 | 0.012815 | 0.1 | 0.88 | |
| 35 | (1, 0, 1, 0, 0, 0, 1, 1) | 0.2 | 0.940504 | 0.005 | 0.012815 | 0.1 | 0.88 | |
| 36 | (1, 1, 1, 0, 0, 0, 1, 1) | 0.8 | 0.940504 | 0.005 | 0.012815 | 0.1 | 0.88 | |
| 37 | (1, 0, 0, 1, 0, 0, 1, 1) | 0.2 | 0.059496 | 0.995 | 0.012815 | 0.1 | 0.88 | 1.34E-05 |
| 38 | (1, 1, 0, 1, 0, 0, 1, 1) | 0.8 | 0.059496 | 0.995 | 0.012815 | 0.1 | 0.88 | 5.34E-05 |
| 39 | (1, 0, 1, 1, 0, 0, 1, 1) | 0.2 | 0.940504 | 0.995 | 0.012815 | 0.1 | 0.88 | 0.000211 |
| 40 | (1, 1, 1, 1, 0, 0, 1, 1) | 0.8 | 0.940504 | 0.995 | 0.012815 | 0.1 | 0.88 | 0.000844 |
| 41 | (1, 0, 0, 0, 1, 0, 1, 1) | 0.2 | 0.059496 | 0.005 | 0.987185 | 0.1 | 0.88 | |
| 42 | (1, 1, 0, 0, 1, 0, 1, 1) | 0.8 | 0.059496 | 0.005 | 0.987185 | 0.1 | 0.88 | 2.07E-05 |
| 43 | (1, 0, 1, 0, 1, 0, 1, 1) | 0.2 | 0.940504 | 0.005 | 0.987185 | 0.1 | 0.88 | |
| 44 | (1, 1, 1, 0, 1, 0, 1, 1) | 0.8 | 0.940504 | 0.005 | 0.987185 | 0.1 | 0.88 | 0.000327 |
| 45 | (1, 0, 0, 1, 1, 0, 1, 1) | 0.2 | 0.059496 | 0.995 | 0.987185 | 0.1 | 0.88 | 0.001029 |
| 46 | (1, 1, 0, 1, 1, 0, 1, 1) | 0.8 | 0.059496 | 0.995 | 0.987185 | 0.1 | 0.88 | 0.004114 |
| 47 | (1, 0, 1, 1, 1, 0, 1, 1) | 0.2 | 0.940504 | 0.995 | 0.987185 | 0.1 | 0.88 | 0.016259 |
| 48 | (1, 1, 1, 1, 1, 0, 1, 1) | 0.8 | 0.940504 | 0.995 | 0.987185 | 0.1 | 0.88 | 0.065036 |
| 49 | (1, 0, 0, 0, 0, 1, 1, 1) | 0.2 | 0.059496 | 0.005 | 0.012815 | 0.9 | 0.88 | |
| 50 | (1, 1, 0, 0, 0, 1, 1, 1) | 0.8 | 0.059496 | 0.005 | 0.012815 | 0.9 | 0.88 | |
| 51 | (1, 0, 1, 0, 0, 1, 1, 1) | 0.2 | 0.940504 | 0.005 | 0.012815 | 0.9 | 0.88 | 9.55E-06 |
| 52 | (1, 1, 1, 0, 0, 1, 1, 1) | 0.8 | 0.940504 | 0.005 | 0.012815 | 0.9 | 0.88 | 3.82E-05 |
| 53 | (1, 0, 0, 1, 0, 1, 1, 1) | 0.2 | 0.059496 | 0.995 | 0.012815 | 0.9 | 0.88 | 0.00012 |
| 54 | (1, 1, 0, 1, 0, 1, 1, 1) | 0.8 | 0.059496 | 0.995 | 0.012815 | 0.9 | 0.88 | 0.000481 |

| | | | | | | | |
|---|---|---|---|---|---|---|---|
| 55 | (1, 0, 1, 1, 0, 1, 1, 1) | 0.2 | 0.940504 | 0.995 | 0.012815 | 0.9 | 0.88 | 0.0019 |
| 56 | (1, 1, 1, 1, 0, 1, 1, 1) | 0.8 | 0.940504 | 0.995 | 0.012815 | 0.9 | 0.88 | 0.007598 |
| 57 | (1, 0, 0, 0, 1, 1, 1, 1) | 0.2 | 0.059496 | 0.005 | 0.987185 | 0.9 | 0.88 | |
| 58 | (1, 1, 0, 0, 1, 1, 1, 1) | 0.8 | 0.059496 | 0.005 | 0.987185 | 0.9 | 0.88 | 0.000186 |
| 59 | (1, 0, 1, 0, 1, 1, 1, 1) | 0.2 | 0.940504 | 0.005 | 0.987185 | 0.9 | 0.88 | 0.000735 |
| 60 | (1, 1, 1, 0, 1, 1, 1, 1) | 0.8 | 0.940504 | 0.005 | 0.987185 | 0.9 | 0.88 | 0.002941 |
| 61 | (1, 0, 0, 1, 1, 1, 1, 1) | 0.2 | 0.059496 | 0.995 | 0.987185 | 0.9 | 0.88 | 0.009257 |
| 62 | (1, 1, 0, 1, 1, 1, 1, 1) | 0.8 | 0.059496 | 0.995 | 0.987185 | 0.9 | 0.88 | 0.037028 |
| 63 | (1, 0, 1, 1, 1, 1, 1, 1) | 0.2 | 0.940504 | 0.995 | 0.987185 | 0.9 | 0.88 | 0.146331 |
| 64 | (1, 1, 1, 1, 1, 1, 1, 1) | 0.8 | 0.940504 | 0.995 | 0.987185 | 0.9 | 0.88 | 0.585325 |
| SUM | | | | | | | | 0.995984 |

Note that in Table 10, if Pr($X_i$) is not empty, $X_i$ is a connected vector. For example, Pr($X_1$) is empty and Pr($X_{64}$) = 0.585325 indicates that $X_1$ is a disconnected vector, and $X_{64}$ is a connected vector.

From the last row in Table 10, we have $R$ = 0.995984, which is the summation of the reliability of all connected vectors in Fig. 5.

## 5. CONCLUSIONS

Various networks are necessary for modern daily life. Reliability of these networks is defined as the success probability that the BN is working, that is, the source and sink nodes are connected by a directed path in the BN. Hence, reliability can be implemented in evaluating, managing, and designing the performance of all types of BNs. However, most reliability problems are focused on the AOA network structure; moreover, they assume that the reliability of each component is known in advance. This paper presents a novel AON BN reliability problem with uncertainty components and presents an AON BAT to solve the problem.

To solve the proposed novel problem, we first invited experts to provide their linguistic variables to rate these uncertainty components. A fuzzy set was then applied. Each linguistic variable was transferred to a fuzzy number, and these fuzzy numbers were aggregated into an AFN corresponding to the same uncertainty component. Each AFN was transferred further to the FPS and FFR, which is considered to an unreliability in the proposed problem.

After transferring each uncertainty component variable to a crisp component, the proposed AON BAT is implemented to calculate the reliability of the AON BNs together with the proposed AON PLSA to verify the connectivity of each obtained vector. Moreover, the time complexity of the proposed AON BAT is $O(2^{n-1})$, with the space complexity being $O(n-1)$ — both of which are less than that of AOA BAT, where $n$ is the number of nodes. In addition, from the example demonstration, it can be observed that the proposed AON BAT can calculate the AON BN reliability with uncertainty components for the 8-node 13-arc 3-uncertainty-component BN. Thus, the proposed algorithm has the potential to solve larger-sized problems.

In future works, the proposed AON BN reliability problem with uncertainty components will be further extended to AON multi-state networks and AON multi-commodity multi-state networks. The proposed algorithm will be improved correspondingly for larger-sized problems.

## ACKNOWLEDGMENT

This research was supported in part by the Ministry of Science and Technology, R.O.C. under grant MOST 107-2221-E-007-072-MY3.